\documentclass{article}
\usepackage{epsfig}
\title{Magneto-photoluminescence studies of
Zn$_{1-x}$Mn$_x$Te/ZnTe multiple quantum-well and quantum dot 
structures}
\author{%
Ivan J. Griffin$^a$, Peter J. Klar$^a$, Daniel Wolverson$^a$%
\thanks{Corresponding author, Fax: +44 1603 259515, 
telephone +44 1603 592980, email: d.wolverson\symbol{64}uea.ac.uk},\\ 
J. John Davies$^a$, Bernard Lunn$^b$, 
Duncan E. Ashenford$^b$,\\
and Torsten Henning$^c$\\[1.5ex]
$^a$School of Physics, University of East Anglia,\\
Norwich NR4 7TJ, UK\\
$^b$Department of Engineering Design and Manufacture,\\
University of Hull, Hull HU6 7RX, UK\\
$^c$Department of Physics, G\"oteborg University and\\
Chalmers University of Technology, SE-41296 G\"oteborg, Sweden}
\date{cond-mat/9805079, 1998-05-07\\
J. Cryst. Growth 184/185 (1998), 325--329}
\begin{document}
\maketitle
\begin{abstract}
Wide quantum dots were fabricated from multiple quantum well 
structures based on Zn$_{1-x}$Mn$_x$Te/ZnTe ($x = 0.076$) dilute magnetic
semiconductors and were investigated via photoluminescence (PL) in a
magnetic field. Calculations taking into account the strain in the two
types of structure enabled the PL transitions to be identified and show
that the dominant emission in the MQWs is from heavy-hole (hh) excitons
whereas in the quantum dots, the removal of the strain in the barrier
layers generates a large biaxial tensile strain in the quantum wells
which shifts the light-hole (lh) exciton to lower energy than the hh
exciton. The lh exciton 
$\sigma^+$ transition is virtually independent of
magnetic field whilst the hh exciton is field-dependent. Thus, at
fields of 1 to 2 Tesla, the hh exciton 
$\sigma^+$ transition again becomes the
lowest energy transition of the quantum dots. These observations are
described by a model with a chemical valence band offset of 30\,\% for
Zn$_{1-x}$Mn$_x$Te/ZnTe.
\end{abstract}

\section{Introduction}
        This paper concerns dots structure fabricated by lithography
from dilute magnetic semiconductor (DMS) heterostructures. Although the
dots do not have sufficiently small radii for lateral quantum
confinement effects to be important, we shall, nevertheless, denote them
by the conventional term 'quantum dots'; the importance of studying such
dots lies in the need to understand strain effects before attempting to
investigate the quantum effects themselves.  In earlier work, we
employed photomodulated reflectivity (PR) and photoluminescence (PL) to
study multiple quantum wells (MQWs) formed from 
Zn$_{1-x}$Mn$_x$Te/ZnTe [1,2]
and to investigate the occurence of a field-induced type I-type II
transition [3]. Dots fabricated from these MQW structures have also been
investigated by PR [4] in the absence of a magnetic field: here we
describe the effects on the PL spectra when such a field is introduced.
\par
\section{Experimental details}
Two Zn$_{1-x}$Mn$_x$Te/ZnTe MQWs were grown by MBE on
(1\,0\,0)-oriented GaSb substrates with 1000\,\AA{} ZnTe buffer layers. Each
heterostructure consisted of 10 ZnTe quantum wells of thickness 80\,\AA{}
(sample number H561) and 100\,\AA{} (H560) embedded between 
Zn$_{1-x}$Mn$_x$Te
barrier layers of $x = 7.6$\,\% and thickness 150\,\AA{}. The samples were
nominally undoped [1]. The dot samples were prepared from pieces of the
two Zn$_{1-x}$Mn$_x$Te/ZnTe MQWs by electron beam lithography for pattern
definition followed by Ar$^{2+}$ ion beam etching for transferring the
pattern, giving a $3.2 \times 3.2$\,mm$^2$ area with a high density of 
homogeneous
tapering pillars of height $\approx 300$\,nm (etching stopped before the
substrate was reached) and diameter of $\approx 200$\,nm [4]. 
PL experiments were
carried out at 1.5\,K in magnetic fields up to 6 Tesla (along the growth
axis). Excitation (433 nm) was provided by a UV-pumped Stilbene 3 dye
laser and the emitted light was detected in the Faraday geometry.
\par

\section{Results and discussion}
\begin{figure}[tbp]
  \centering
    \epsfig{file=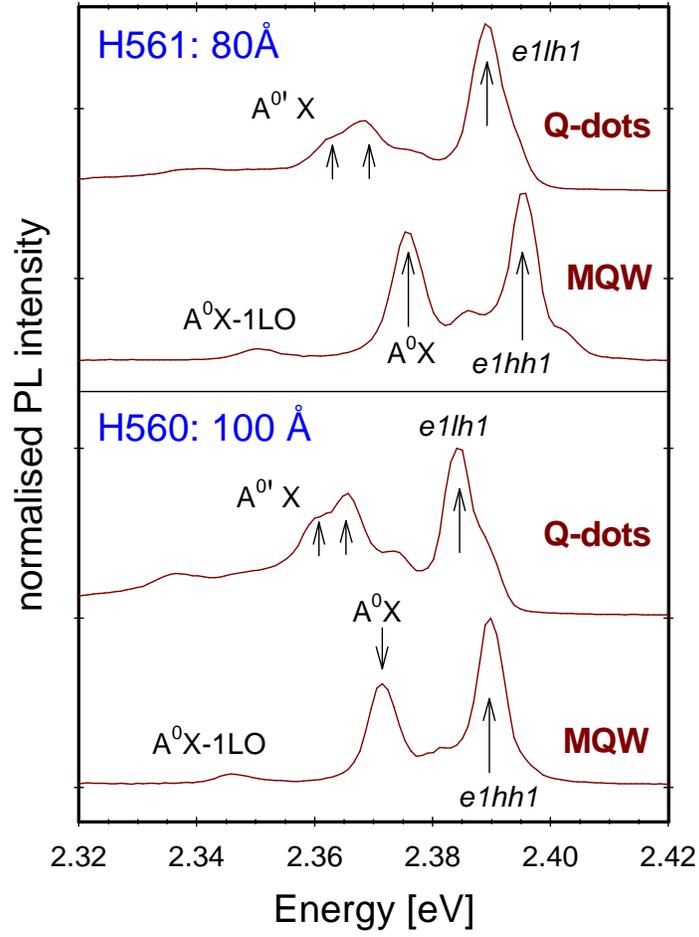,width=0.8\textwidth}
    \caption{Photoluminescence spectra at 1.5\,K with 433\,nm 
      excitation of the two 
      multiple Zn$_{1-x}$Mn$_x$Te/ZnTe quantum well samples (well widths 
      indicated in the figure) and of the quantum dot 
      structures fabricated from them. The spectra are normalised to the 
      same peak height for the 
      dominant transition in each case. 
      The labels of the various peaks are explained in the text.}
    \label{fig:fig1}
\end{figure}
Figure~1 shows the zero-magnetic field PL spectra
of the two Zn$_{1-x}$Mn$_x$Te/ZnTe MQW samples and the two corresponding dot
samples. We first discuss the free excitons region (above about 2.38\,eV). 
Fig.~1 shows, firstly, that the PL emission from the quantum wells in
the dot samples (labelled 
\textit{e1lh1}) is shifted to lower energy by about
6\,meV 
compared to the corresponding bands (\textit{e1hh1}) in the PL of the
MQWs. Secondly, because of stronger confinement in narrower quantum
wells, the free exciton emission for the MQW and dot from H561 is at a
higher energy than for H560.
\par

The first observation is of particular
interest here since it relates to differences in strain.  To a first
approximation, the MQW samples are strained to the ZnTe lattice
constant, resulting in a compressive biaxial strain in the barriers and
zero strain in the quantum wells [1]. In contrast, the dots appear to be
free-standing structures with an interatomic spacing in the layer plane
intermediate between that of Zn$_{1-x}$Mn$_x$Te and ZnTe, there being a
consequent tensile biaxial strain in the quantum well layers [2]. For
the two cases there is a different ordering of the excitonic energy
levels: in the original, unstrained quantum wells, the free exciton
emission is expected to have heavy-hole character (confinement effects
are smaller for the heavy holes, so that the 
\textit{e1hh1} is the transition of
lowest energy); in contrast, for the dots, there will exist a tensile
biaxial strain which may be large enough for the light hole states to
become lowest. The experiments described below confirm that this is so.
\par

In the bound state PL (below about 2.38\,eV), an acceptor-bound exciton
line (A$^0$X) and its first phonon replica (A$^0$X-1LO) can be seen for both
the original MQW structures (Fig. 1).  This acceptor is presumed to
originate from Sb diffused from the GaSb substrate [5]. The A$^0$X signal
shows no significant magnetic field dependence (see below) and is
attributed to the buffer layer (in which the Sb content will presumably
be highest). In the PL spectra of the dots, the A$^0$X band is weaker and
an additional acceptor band, A$^{0^\prime}$X, is observed, 
probably resulting from
the nanofabrication process, as reported by other groups [6,7].\par

\begin{figure}[tbp]
  \centering
  \epsfig{file=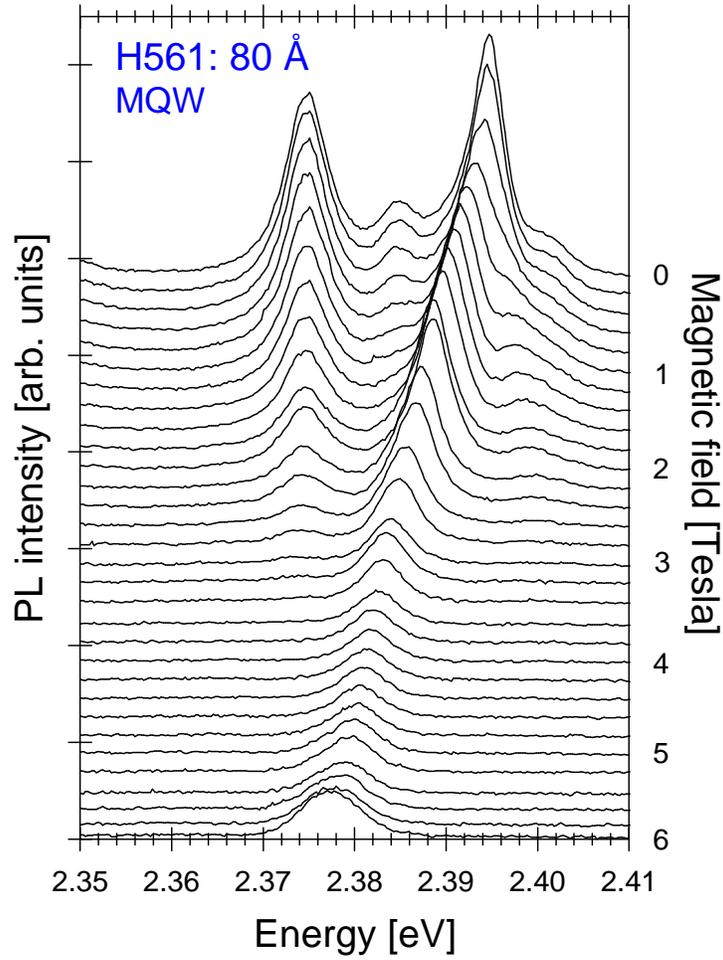,width=0.8\textwidth}
  \caption{Photoluminescence spectra of the multiple quantum well 
    sample of well width 80\,\AA{} at 1.5\,K 
    and in magnetic fields from 0 to 6~Tesla. 
    The spectra are displaced vertically downwards in 
    proportion to the magnetic field and are not normalised. 
    For the identification of the bands at 
    zero magnetic field, see Fig.~1.}
  \label{fig:fig2}
\end{figure}
\begin{figure}[tbp]
  \centering
  \epsfig{file=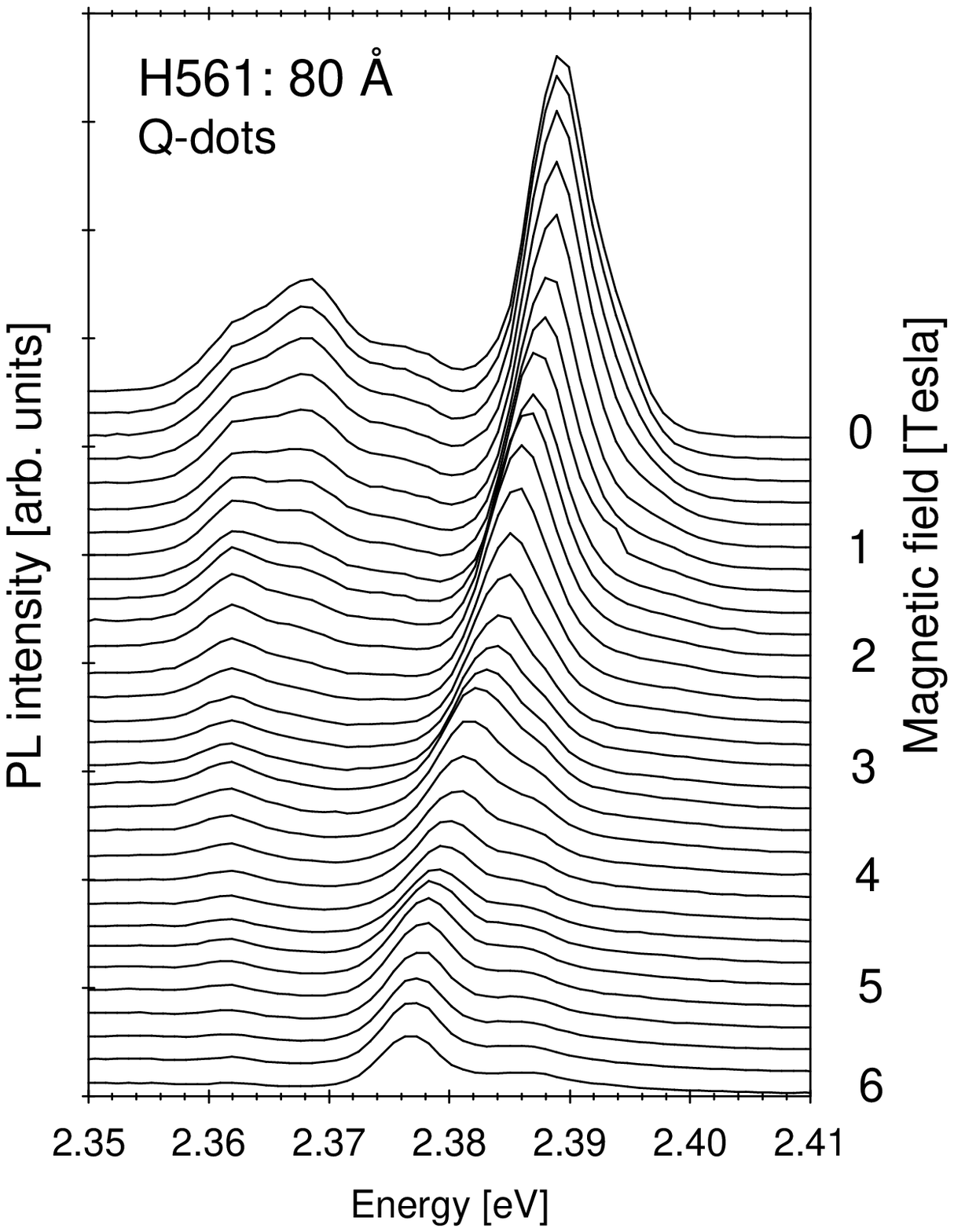,width=0.8\textwidth}
  \caption{Photoluminescence spectra of the quantum dots fabricated 
    from the multiple quantum well 
    sample of well width 80\,\AA{} at 1.5\,K and in magnetic 
    fields from 0 to 6~Tesla. The spectra are 
    displaced vertically downwards in proportion to the 
    magnetic field and are not normalised. 
    For the identification of the bands at zero magnetic 
    field, see Fig.~1.
}
  \label{fig:fig3}
\end{figure}
The
effects of a magnetic field on the PL spectra are shown in Figs.~2 and
3 for the MQW and the dots of H561. The quantum well-related emission
bands shift to lower energy with increasing field, whilst the signals
attributed to the buffer layer show no discernible shift.  We note that
the excitonic PL transition 
$\sigma^+$~\textit{e1hh1} (involving the 
$j_z = -3/2$ heavy hole
and the $s_z = -1/2$ electron states) is expected to shift more rapidly
with increasing magnetic field than the light-hole like PL transition 
$\sigma^+$~e1lh1 
($j_z = -1/2$, $s_z = +1/2$); this is because, firstly, the
field-induced splitting of the heavy hole states is approximately three
times bigger than that of the light hole states and, secondly, the
magnetic shifts of the single-particle hole and electron states
constituting the exciton \emph{add} for the \textit{hh} exciton but have 
\emph{opposite} sign
for the 
\textit{lh} exciton and approximately cancel. The character of the free
exciton emission should therefore change from light to heavy hole-like
with increasing field: we shall show that this does indeed occur.\par

The
intensities of the quantum well-related bands (Figs.~2 and 3) change with
magnetic field, but this behaviour is hard to interpret, since the
magnetic field induces simultaneously changes in the occupation of the
states, in the non-excitonic recombination processes and in the degree
of confinement. The occurence of a magnetic-field induced type I-type II
transition for the $\sigma^+$ e1lh1 exciton [3,8] further complicates the
situation and we therefore concentrate on the energy changes, rather
than the intensity changes, in the PL spectra.  Full details of the
calculations of the exciton energies are given elsewhere [3] but can be
summarised as follows.  Five separate steps are carried out for each
magnetic field value and for both polarisation states. First, the
single-particle potentials for the electrons and the holes are
calculated. Strain shifts, interface roughness effects and magnetic
field shifts may be included at this stage (we used the model of Stirner
et al. [3,9] and assumed ideally smooth interfaces; enhanced
paramagnetism effects, which arise because Mn$^{2+}$ ions in the barrier near
to the interface to the quantum well have a lower probability of
antiferromagnetic spin-pairing with neighbouring Mn$^{2+}$ ions than do those
deep in the barrier, have not been included since such effects assume
greatest importance when the Mn$^{2+}$ content of the barrier is large 
($> 10$\,\%) [9]). The remaining steps use a method based on that of Peyla et
al. [10] to calculate the excitonic transition energies within the
potentials obtained from the first step. The electron single- particle
problem is solved and an effective potential for the holes is calculated
which includes the QW potential plus that due to the confined
electron. The hole energies (and thus the transition energies) are found
and an iterative approach used to minimise the exciton energy by
variation of the exciton radius parameter in the trial wave
function.\par

\begin{figure}[tbp]
  \centering
  \epsfig{file=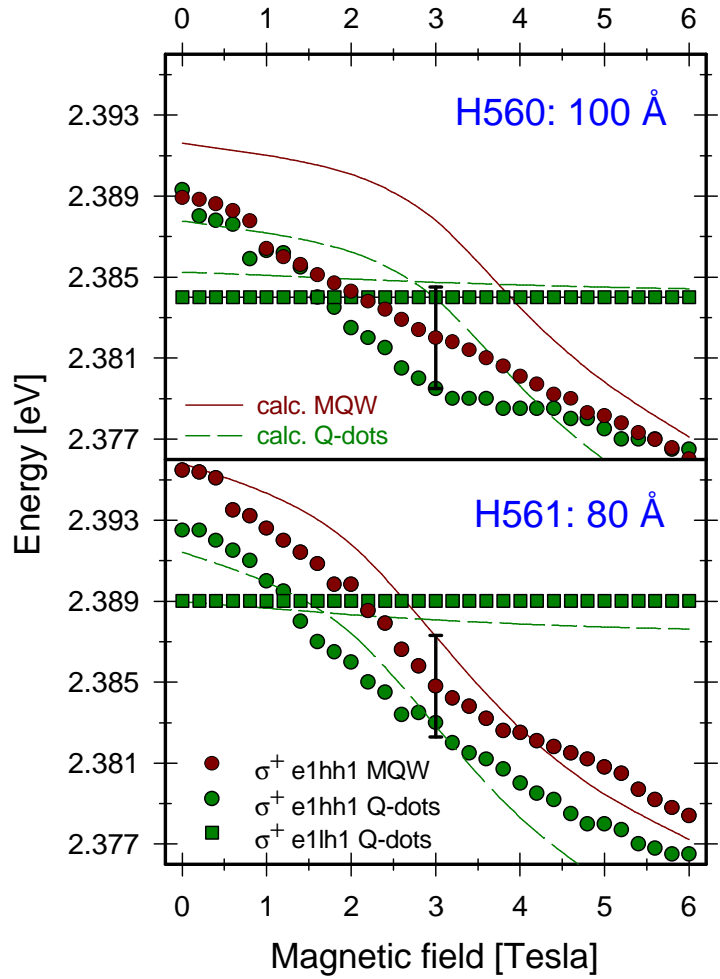,width=0.8\textwidth}
  \caption{The measured and calculated transition energies of the 
    quantum well excitonic 
    photoluminescence bands as a function of magnetic field 
    for the four structures studied. Solid 
    symbols and solid lines: experimental data and calculations 
    respectively for the multiple 
    quantum wells. Open symbols and dashed lines: experimental 
    data and calculations 
    respectively for the quantum dot structures. The vertical 
    bars indicate the linewidths of the PL 
    bands.}
  \label{fig:fig4}
\end{figure}
The results of these calculations are shown in Fig.~4. We consider
first the PL from the MQWs (only the heavy-hole transitions are
observed). Using a chemical valence band offset of 30\,\% of the
unstrained band gap difference, a barrier Mn concentration of 0.076 and
well widths of 100\,\AA{} and 80\,\AA{} respectively with the assumption of
smooth interfaces, the calculation yields the solid lines shown in
Fig.~4. The agreement with experiment (solid symbols) is good for H561
and poorer for H560: however, the set of parameters used is that which
gives the best overall description of the 
\emph{complete} set of excitonic
transitions (both $\sigma^+$ and $\sigma^-$) in the barriers and 
wells as observed by
modulated photoreflectivity [3].\par

The same calculation can be applied to
the dots, since these are wide enough that no new lateral confinement
effects are anticipated. The strain state is modified in the way
discussed earlier and leads to the calculated values shown in Fig. 4
(dashed lines). Comparison with experiment shows that, again, the
agreement is good for H561 and poorer for H560.  However, the main
prediction (that in the dots the character of the PL should change from
lh- like to hh-like as the magnetic field increases) is borne out in
both cases. The near-independence of the lh transition on magnetic
field is very clear in Fig.~4 , which shows the crossing of the 
\textit{lh} and
\textit{hh} hole states at a magnetic field of around 1~Tesla.\par

\section{Conclusions}
The data obtained from photoreflectivity studies of parent MQW layers
were used to predict the excitonic PL energies in dots fabricated from
these structures. The only parameter that was changed was the strain: in
the dots, the quantum well regions become subject to a large biaxial
tensile strain, which causes the lh exciton state to lie below that of
the heavy hole, a situation that can then be reversed by application of
a magnetic field. The results demonstrate the usefulness of DMS
materials in investigations of low-dimensional structures in general and
highlight the importance of first establishing the effects of changes in
strain before reducing dot diameters sufficiently for lateral quantum
confinement to occur.  

\section*{Acknowledgements}
IJG and PJK thank EPSRC (UK) and
UEA-Norwich respectively for research studentships.  The work was
supported by EPSRC under grants GR/H57356, GR/H93774 and GR/K04589.  We
are also grateful to the Swedish Nanometer Laboratory for the use of
their facilities.


\begin{thebibliography}{99}
\itemsep0.5\itemsep
\bibitem[1]{1}     P. J. Klar, C. M. Townsley, D. Wolverson, J. J. Davies, 
D. E. Ashenford and B. Lunn, 
Semicond. Sci. Technol. 10 (1995) 1568
\bibitem[2]{2}     P. J. Klar, D. Wolverson, D. E. Ashenford and B. Lunn, 
J. Cryst. Growth 159 (1996) 
528
\bibitem[3]{3}    P. J. Klar, J. R. Watling, D. Wolverson, J. J. Davies, 
D. E. Ashenford and B. Lunn, 
Semicond. Sci. Tech., (1997) in press 
\bibitem[4]{4}    P. J. Klar, D. Wolverson, D. E. Ashenford,  B. Lunn and 
T. Henning, Semicond. Sci. 
Technol. 11 (1996) 1863
\bibitem[5]{5}     N. J. Duddles, J. E. Nicholls, T. J. Gregory, 
W. E. Hagston, B. Lunn and D. E. 
Ashenford, J. Vac. Sci. Technol. B 10 (1992) 912
\bibitem[6]{6}    M. Illing, G. Bacher, A. Forchel, A. Waag, T. Litz 
and G. Landwehr, J. Crystal 
Growth 138 (1994) 638
\bibitem[7]{7}    C. Gourgon, L. S. Dang, H. Mariette, C. Vieu, and 
F. Muller, Appl. Phys. Letts. 66 
(1995) 1635
\bibitem[8]{8}     H. H. Cheng, R. J. Nicholas, M. J. Lawless, 
D. E. Ashenford and B. Lunn, 
Phys. Rev. B 52 (1995) 5269
\bibitem[9]{9}     T. Stirner, J. M. Fatah, R. J. Roberts, T. Piorek, 
W. E. Hagston and P. Harrison, 
Superlatt. Microstruct. 16 (1994) 11
\bibitem[10]{10}    P. Peyla, Y. Merle d'AubignÈ, A. Wasiela, 
R. Romestain, H. Mariette, M. D. Surge, N. 
Magnea and H. Tuffigo, Phys. Rev. B 46 (1992) 1557
\end{thebibliography}
\end{document}